# Galvanic vestibular stimulation produces cross-modal improvements in visual thresholds

Jamie L. Voros (1), Sage O. Sherman (1), Rachel Rise (1), Alexander Kryuchkov (1), Ponder Stine (1), Allison P. Anderson (1), and Torin K. Clark (1)

((1) Ann & H.J. Smead Department of Aerospace Engineering Sciences, University of Colorado-Boulder)

Background: Stochastic resonance (SR) refers to a faint signal being enhanced with the addition of white noise. Previous studies have found that vestibular perceptual thresholds are lowered with noisy galvanic vestibular stimulation (i.e., "in-channel" SR). Auditory white noise has been shown to improve tactile and visual thresholds, suggesting "cross-modal" SR.

Objective: We aimed to study the cross-modal impact of noisy galvanic vestibular stimulation (nGVS) (n=9 subjects) on visual and auditory thresholds.

Methods: We measured auditory and visual perceptual thresholds of human subjects across a swath of different nGVS levels in order to determine if a subject-specific best nGVS level elicited a reduction in thresholds as compared the no noise condition (sham).

Results: We found an 18% improvement in visual thresholds (p = 0.026). Among the 7 of 9 subjects with reduced thresholds, the average improvement was 26%. Subjects with higher (worse) visual thresholds with no stimulation (sham) improved more than those with lower thresholds (p = 0.005). Auditory thresholds were unchanged by vestibular stimulation.

Conclusions: These results are the first demonstration of cross-modal improvement with nGVS, indicating galvanic vestibular white noise can produce cross-modal improvements in some sensory channels, but not all.

## Introduction

Stochastic resonance (SR) is a phenomenon whereby an input signal to a non-linear system is enhanced by the presence of a particular non-zero level of noise [1]. SR in human physiological sensory systems has been observed, in which a faint signal (stimulus) is perceived more easily with the addition of white noise [1]–[5]. In-channel SR refers to stochastic resonance occurring within the same sensory modality (e.g. auditory white noise improving auditory perception). Cross-modal SR refers to stochastic resonance occurring outside the sensory modality of the white noise (e.g. vestibular white noise improving visual perception).

SR has often been investigated and observed though psychophysiological experiments, aimed at quantifying perceptual thresholds [6]–[11]. A perceptual threshold is the smallest stimulus input that can still be reliably perceived by a person. For example, an auditory threshold refers to the faintest sound one can still reliably hear. In the domain of perceptual thresholds, SR is thought to show a characteristic u shape of as a function of white noise as shown in Figure 1 [2], [3], [8], [12].

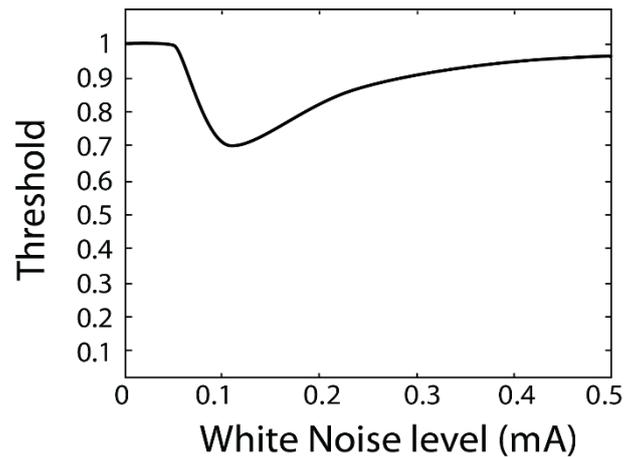

*Figure 1 Graph to show characteristic shape of SR the curve in threshold against noise level.*

Specifically, as more white noise is added it is thought to resonant with the stimulus to produce a reduced perceptual threshold, but when too much white noise is added it is no longer beneficial, and for some in-channel sensing modalities can degrade perception. SR in the visual channel (typically as white noise added to images) is a well-documented occurrence in subjects with healthy vision [13], [14], [14]–[16] and has also been demonstrated in visually impaired subjects [17]. Additionally, auditory white noise has been shown to lower auditory thresholds in subjects with healthy hearing [10], [11], [18] and those with cochlear implants [11]. White noise in the same sensory channel has been found to improve touch [20]–[23] and vestibular perceptual thresholds [6], [7], [24], [25] as well as functional vestibular responses such as balance in the dark, spinal reflexes, and locomotion [24-28].

Vestibular perception may be altered by applying electrical white noise via electrodes placed on the mastoids, referred to as galvanic vestibular stimulation (GVS) [6], [7], [24], [26]. Improvements in roll tilt vestibular thresholds exist within the subject pool but are not consistent from subject to subject, ranging from a 50% reduction in threshold to no improvement at all [6], [7]. There are also inconsistencies as to the electric current level of noisy GVS (nGVS) eliciting an improvement in vestibular thresholds [6], [7]. Vestibular stimulation in healthy subjects appears to only produce benefits during active stimulation [27], while others have suggested improved balance in elderly patients even after stimulation has ceased [28].

Each of the aforementioned studies applied white noise to the same channel in which perception was measured, but cross-modal SR has also been demonstrated. Cross-modal SR is achieved when improvements in perception occur in a different channel from that of the white noise stimulation [8], [9]. Previous studies have suggested that applying auditory white noise can improve visual flicker sensitivity [9], [29], visual contrast thresholds [8] and motor control [30]. We note relevant caveats to these studies: The first study [9] statistically compares sham thresholds to each whatever noise level happened to produce each individual subject's best threshold. This post-hoc selection without an independent reassessment will produce a biased sample and increase the likelihood of a false positive. The third [8] does not statistically assess findings, but demonstrates descriptive improvements. The second and fourth [29], [30] use data from just three and four subjects respectively. All studies support the notion that there is not one white noise level that is optimal for all subjects, as each subject had an individually-determined optimal stimulation level [8], [9], [29], [30]. Another study showed tactile stimulation to enhance speech recognition in subjects with cochlear ear implants [31], which was later hypothesized to be due to the multisensory nature of the dorsal cochlear nucleus [32]. We are not aware of any studies investigating cross-modal SR by applying white noise to the vestibular system using GVS.

In this research, we aimed to test for the presence of cross-modal SR in auditory and visual sensory modalities with the application of nGVS. We built upon observations of in-channel SR in auditory and visual modalities and the previously investigated cross-modal benefits of auditory white noise. Instead of auditory white noise, here we explored using GVS owing to its efficacy in improving vestibular thresholds and balance. Since many studies have demonstrated optimal noise levels to achieve SR are individualized [6], [8], [9], our methods ensure independent samples between thresholds measured with nGVS and thresholds measured without nGVS (sham). By first determining the best nGVS level (for each subject), we were able to then re-measure the subjects' threshold with no stimulation (sham) and with the best nGVS level for two independent, randomized samples for a paired statistical test.

# Method

## Subjects

Ten unique subjects were enrolled and passed the screening criteria described below (4F, ages 18-25 mean 21.4 years). Eight subjects completed all testing for both visual and auditory threshold tasks, one subject completed only the visual task and one other subject did not do the re-measure (see SR detection) protocol in the visual task.

All subjects were healthy with no known history of vestibular dysfunction, hearing difficulty, tactile dysfunction or vision that could not be corrected with contact lenses. Three potential subjects were removed due to requiring glasses (and not contact lenses) in order to have normal vision, which were not compatible with our testing apparatus. All procedures were approved by the University of Colorado-Boulder Institutional Review Board and all subjects provided written informed consent.

## Study Design

After screening, subjects returned to the laboratory on two subsequent visits (separate days within a two-week period) to complete testing. One visit tested all visual thresholds and the other all auditory thresholds. The GVS electrodes were (re)applied (see Vestibular White Noise Application section below) and removed at the beginning and end of each testing visit.

The GVS system was donned prior to any testing and worn for the remainder of the visit (including during sham condition), however galvanic stimulation was only applied during threshold measurement sessions. Subjects were provided a several minute break between sessions, but the electrodes were not removed. Galvanic vestibular white noise was applied bilaterally via electrodes placed on the mastoids. Broadband (0-100kHz), unipolar, zero-mean white noise was generated by the stimulator (Soterix Medical Inc., Model 0810) and delivered via leads connected to electrodes with a total contact area of 2cm^2. The surface of the skin was prepared with Nuprep skin prep gel and cleaned with alcohol wipes. Electrodes were then placed, secured with a headband, and then Signagel electrode gel (Parker Labs) was injected to the electrode sites. Stimulation was applied only after impedance was indicated as acceptably low by an indicator on the device. The magnitude level of the white noise stimulation was defined as the peak current level.

Thresholds (either visual or auditory, see next section) were assessed over a range of nGVS current levels from 0 mA to 1 mA in increments of 0.1 mA in a randomized order. The subject-specific nGVS level which yielded the best perception (i.e., their 'best' nGVS level or bnGVS) was defined as the white noise level (not including sham) resulting in the lowest measured threshold. The subjects' perceptual thresholds at the sham and bnGVS noise levels were then re-measured to generate independent samples. The order in which the re-measured sham threshold and threshold at bnGVS level were tested was also randomized. The bnGVS level was determined independently for auditory and visual thresholds, such that a given subject often had different bnGVS levels for the two threshold modalities.

All threshold measurements were performed inside a darkroom and sound booth to minimize sensory cues outside the modality in which the threshold was being measured. Subjects and test operators were blinded to the stimulation condition. It is possible that at the highest stimulation levels some subjects could have felt a tingling sensation, but they were not primed to know this would have meant higher levels of GVS stimulation.

### Perceptual thresholds

Thresholds were measured with a two-alternative forced-choice detection task, in which that subject had to identify which of two sequential intervals the stimulus was in. The stimulus (e.g., auditory tone) always occurred in either the first or second interval, with no stimulus (e.g., no auditory tone) occurring in the other, determined randomly for each trial. Subjects responded verbally (e.g. "interval one" or "interval two") to indicate which interval they thought contained the stimulus. An adaptive 3 down 1 up Parametric Estimation by Sequential Testing (PEST) [33]–[35] procedure was used to determine the magnitude of the stimuli (e.g., loudness of the auditory tone) for each trial. Subject responses were fit with a cumulative Gaussian psychometric function [33], [36], [37] scaled from 0.5 to 1 (since guessing performance would yield 0.5 percent correct with the two alternatives). The cumulative Gaussian was parameterized by two values, $\mu$ and $\sigma$. Here, the $\mu$ value represented the stimulus level at which the subject stands to get 75% of trials correct, which we defined as the threshold.

The threshold estimation theoretically becomes more precise with more trials to which the psychometric curve can be fit. However, subject fatigue, focus, and availability can practically constrain this benefit. Informed by performing Monte-Carlo simulations [38] alongside pilot studies, we chose to perform 50 trials for each visual threshold test (at a given white noise level) and 100 trials for each auditory threshold test. Similarly, re-measures had 50 trials at each of sham and bnGVS for visual thresholds and 100 for auditory. Pilot testing suggested visual thresholds could be estimated well with fewer trials due to typically steeper slopes (lower $\sigma$ values) in the psychometric curves observed, which enabled more efficient estimation of the $\mu$ parameter (threshold).

We used contrast gratings to measure visual contrast thresholds [39]. In each 1 second interval, subjects were presented with one of the types of patches shown in Figure 2. Subjects had to identify which interval contained the patch with the grating. Each visual grating (Figure 2) was 21 cm tall and wide (square) and was presented on an otherwise grey computer monitor placed 30 cm in front of the seated subject near eye level.

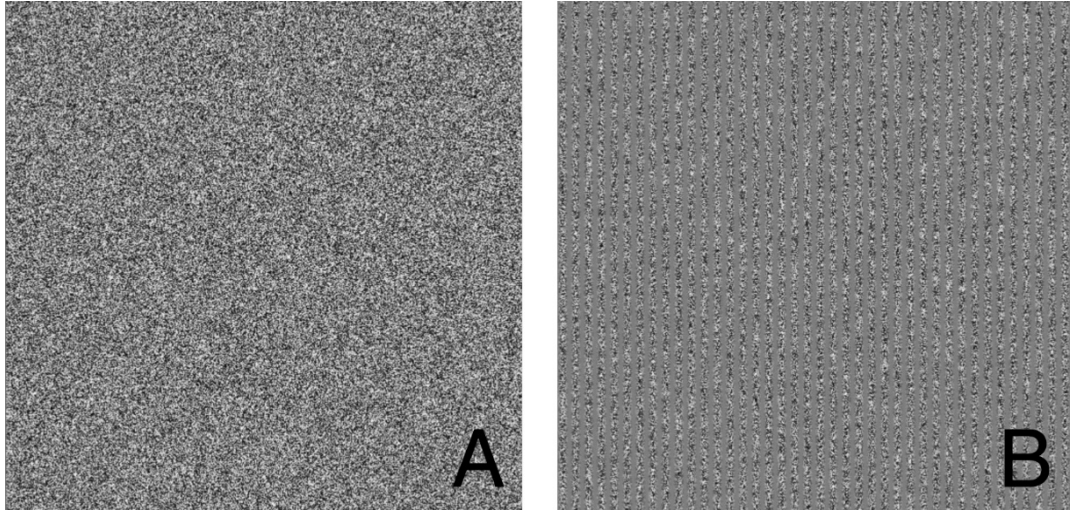

*Figure 2: Visual threshold task example presentations. Panel A: Patch containing only visual static noise (i.e., no signal). Panel B: Patch containing 40 vertical gratings (i.e., signal). Subjects were tasked with determining which interval presentation (first or second) contained the vertical gratings.*

Auditory thresholds were measured in the right ear with a 1 kHz pure tone stimulus of 0.25 seconds in duration. Subjects were presented sequentially with two 0.25 second intervals, separate by another 0.25 seconds, in which one (and only one) interval contained the auditory tone. Subjects had to identify which interval contained the tone. Auditory tones were administered via a device (Creare Hearing Assessment, Creare Inc.) and though over-the-ear headphones.

## Analysis

To assess our hypothesis that GVS improved thresholds compared to sham, a one tailed t-test was performed between the re-measured sham thresholds and re-measured thresholds with stimulation. The Shapiro-Wilk test for normality was performed on the paired differences to ensure normal distribution of visual and auditory thresholds.

In order to detect the characteristic U shape associated with SR, we used a subjective human classification method previously described [6], [33], [38], [40]. Briefly, judges viewed plots of measured threshold versus nGVS level, similar to those shown in Figure 3A. Judges were given plots of actual subject data randomly interspersed with plots from simulated subjects and were asked to classify each plot as exhibiting SR (via the characteristic U shape) or not exhibiting SR (expected no U dip). Simulated subjects were modelled with the same experimental protocol of real subjects (e.g., number of trials, adaptive sampling, psychometric curve fitting) [6], [33], [38], [40]. Simulated subjects had a 50% split of not having underlying SR (constant underlying threshold at each nGVS level) or having underlying SR (we assumed an underlying threshold reduction of 30% at the minimum of the U shape, motivated by that previously observed [6]). Critically, the measured thresholds include measurement variability due to the finite number of trials, such that classifying each plot as exhibiting SR was non-trivial (as it is with experimental subject data). Two human judges classified 90 simulated subjects along with 10 subjects for visual thresholds and 9 subjects for auditory thresholds (recall that of the nine subjects who completed the visual thresholds, one did not return to complete the auditory thresholds). Both judges were authors and were familiar with SR curve shape, but they were blinded as to whether each plot was simulated or an experimental subject. Judging classifications were assessed via chi-squared tests for differences between pairs of each of the three groups: simulated subjects with SR, simulated subjects without SR and actual subjects. For example, a chi-squared test between actual subject classifications and simulated subjects with SR classifications can indicate if the proportion of plots the judges classified as having SR differed between the two groups.

# Results

Figure 3 shows an example subject's data. For the visual thresholds (Figure 3A) with nGVS of 0.1mA the threshold was reduced (i.e., improved) relative to the sham threshold. Further increases of nGVS caused the thresholds to increase to near or above the sham threshold. The re-measure thresholds (shown as circles) performed at the bnGVS level of 0.1mA and sham, also showed a lower threshold at 0.1mA as compared to sham. The auditory thresholds for this same subject (Figure 3B) were fairly consistent for each level of nGVS tested. The bnGVS level was identified as 0.3mA, but re-measuring the threshold with bnGVS yielded minimal improvement over the re-measured sham

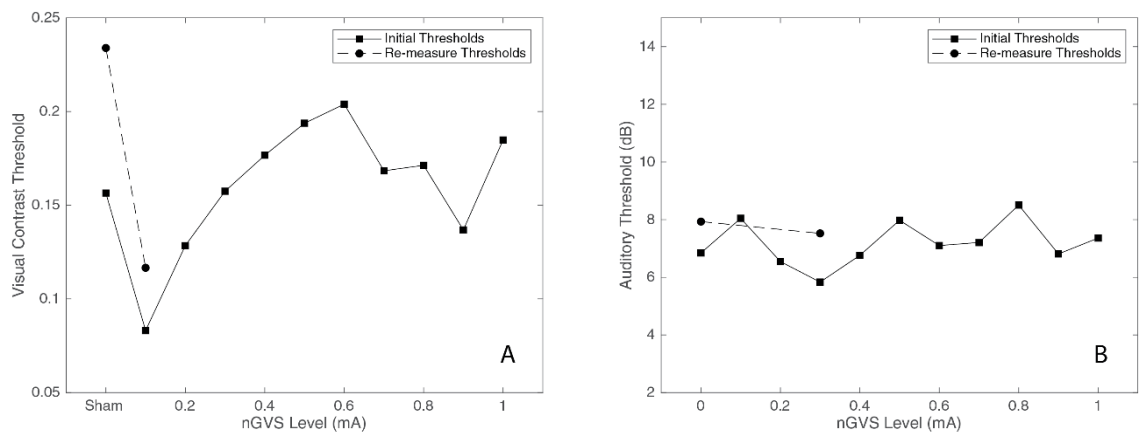

*Figure 3 Plots of threshold against nGVS level for one example subject. Left: Visual contrast threshold measurements, bnGVS is at 0.1mA. Right: Auditory threshold measurements, bnGVS is at 0.3mA.*

## bnGVS Levels

Similar to previous studies [6], [7], we found considerable variation across subjects in the nGVS level resulting in the lowest measured threshold (i.e., the best nGVS level, bnGVS). Figure 4 shows histograms of bnGVS level split by task (visual, auditory). The bnGVS level for both tasks varied across the full range we tested, from 0.1mA to 1mA in intervals of 0.1mA. Further, Figure 4C shows each subject's bnGVS for the visual versus that for the auditory tasks; no correlation was observed (Pearson correlation $r(7) = 0.11$, $p = 0.77$), thus showing the best nGVS level was not consistent for an individual between the visual and auditory tasks.

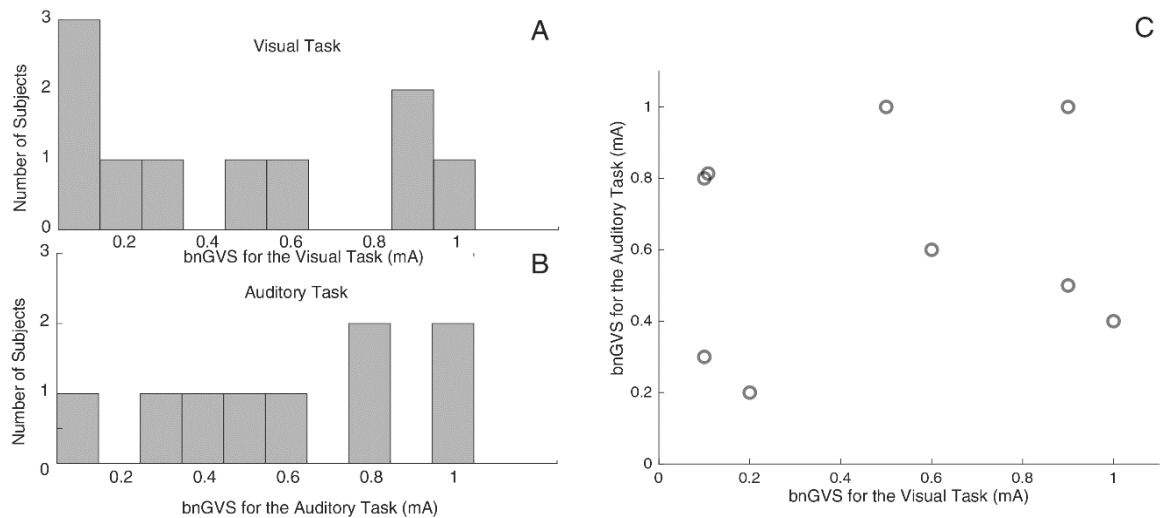

*Figure 4 Histograms to show GVS levels resulting in the lowest threshold measurement. Panel A: Visual task bnGVS, panel B: Auditory task bnGVS. The visual task had ten subjects complete testing and the auditory task had just nine. Panel C: Each subject's Visual task bnGVS level versus their Auditory task bnGVS, showing the nGVS that produced the lowest threshold in each task were unrelated.*

## Indicators of SR

In order to discern a difference in perception with the addition of white noise, we performed independent re-measures of the sham and that which was determined to be the best GVS white noise level.

As our primary finding, the visual thresholds (Figure 5A) were statistically significantly lower in the re-measure with the subject-specific best GVS white noise than in the re-measure with sham (paired t-test, $t(8) = 2.27$, mean difference = -0.038, $p = 0.026$, 95% confidence interval (CI) = [-∞, -0.007]). The mean improvement of 0.038 corresponds to an 18% improvement relative to the mean sham threshold. Among just the seven (of nine) subjects that had benefits from the GVS white noise, the improvement averaged 0.056, a 26% improvement relative to the average sham threshold.

For the auditory thresholds (Figure 5B), there was no significant difference found between the sham and best re-measures (paired t-test, $t(8) = 0.188$, mean difference = -0.16 dB, $p = 0.43$, 95% CI = [-1.4, ∞]). While most subjects did have slight improvements (i.e., lower thresholds) in the re-measure with bnGVS, several subjects actually had worse thresholds with nGVS.

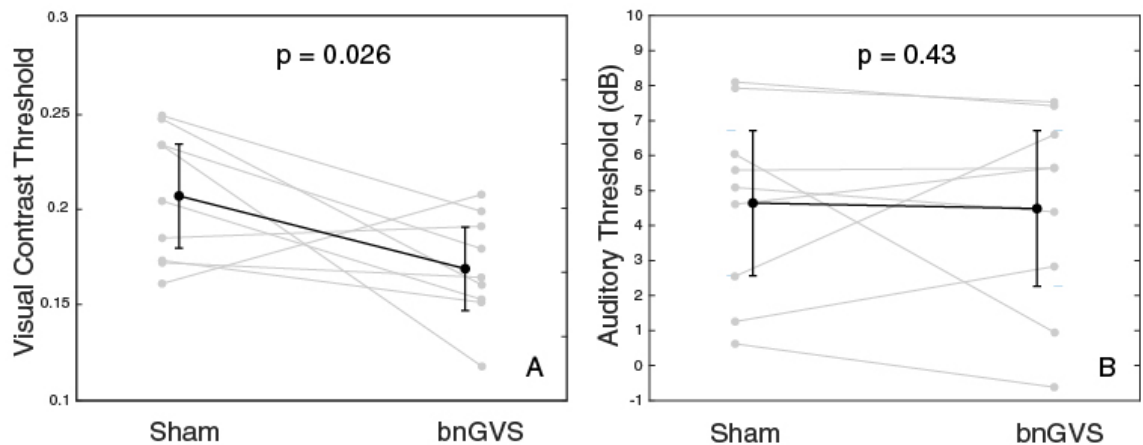

*Figure 5 Plots to show visual (Panel A) and auditory (Panel B) thresholds with and without GVS. Visual thresholds were statistically significant improved with bnGVS.*

In order to determine whether SR was the underlying mechanism responsible for threshold improvement [6], [38], we had blinded judges classify whether simulated and real subject datasets exhibited SR (Figure 6). While some of our experimental subjects were classified as having SR, most were not (rightmost bar in each panel). This tended to contrast the simulations which had underlying SR, which were predominantly classified (correctly) as exhibiting SR. Critically, the simulated subjects with no underlying SR were occasionally misclassified as having SR (i.e., a false positive). This highlights the importance of comparing experimental subject outcomes to those simulated with no underlying SR to properly account for false positives.

Judge #1 on the visual task (Figure 6A), classified experimental subjects differently from simulated subjects exhibiting SR ($\chi^2$(DOF = 1, N = 54)=14.8, p < 0.001) but not differently from simulated subjects exhibiting no SR ($\chi^2$(DOF = 1, N = 55) = 2.5, p = 0.11). Judge #2 for the visual task (Figure 6B), did not differentiate between simulations with and without SR as well as judge #1. By judge #2's classifications, the subject pool was not significantly different from either simulation group: simulations with SR as compared to subjects ($\chi^2$(DOF = 1, N = 54) = 2.3, p = 0.13) and simulations without SR as compared to subjects ($\chi^2$(DOF = 1, N = 55) = 0.27, p = 0.60). While Judge #2 was inconclusive, Judge #1's classifications suggest that our subjects' visual thresholds did not demonstrate the characteristic u-shaped SR curve.

For the auditory task, judge #1's subject classifications (Figure 6C) were different from both simulations with SR ($\chi^2$(DOF = 1, N = 60) = 26, p < 0.001) and without SR ($\chi^2$(DOF = 1, N = 48) = 4.4, p = 0.036). Judge #2's classifications of subjects were different from simulations with SR ($\chi^2$(DOF = 1, N = 60) = 8.9, p = 0.003) and consistent with simulations without SR ($\chi^2$(DOF = 1, N = 48) = 0.11, p = 0.74). While Judge #1's classifications suggest that the subject pool lies somewhere between simulations with SR and simulations without SR, Judge #2's classifications imply that the subject group is most consistent with simulations without SR. Thus, this blind-judging classification analysis suggests nGVS does not produce the characteristic u-shaped SR curve in either visual or auditory thresholds.

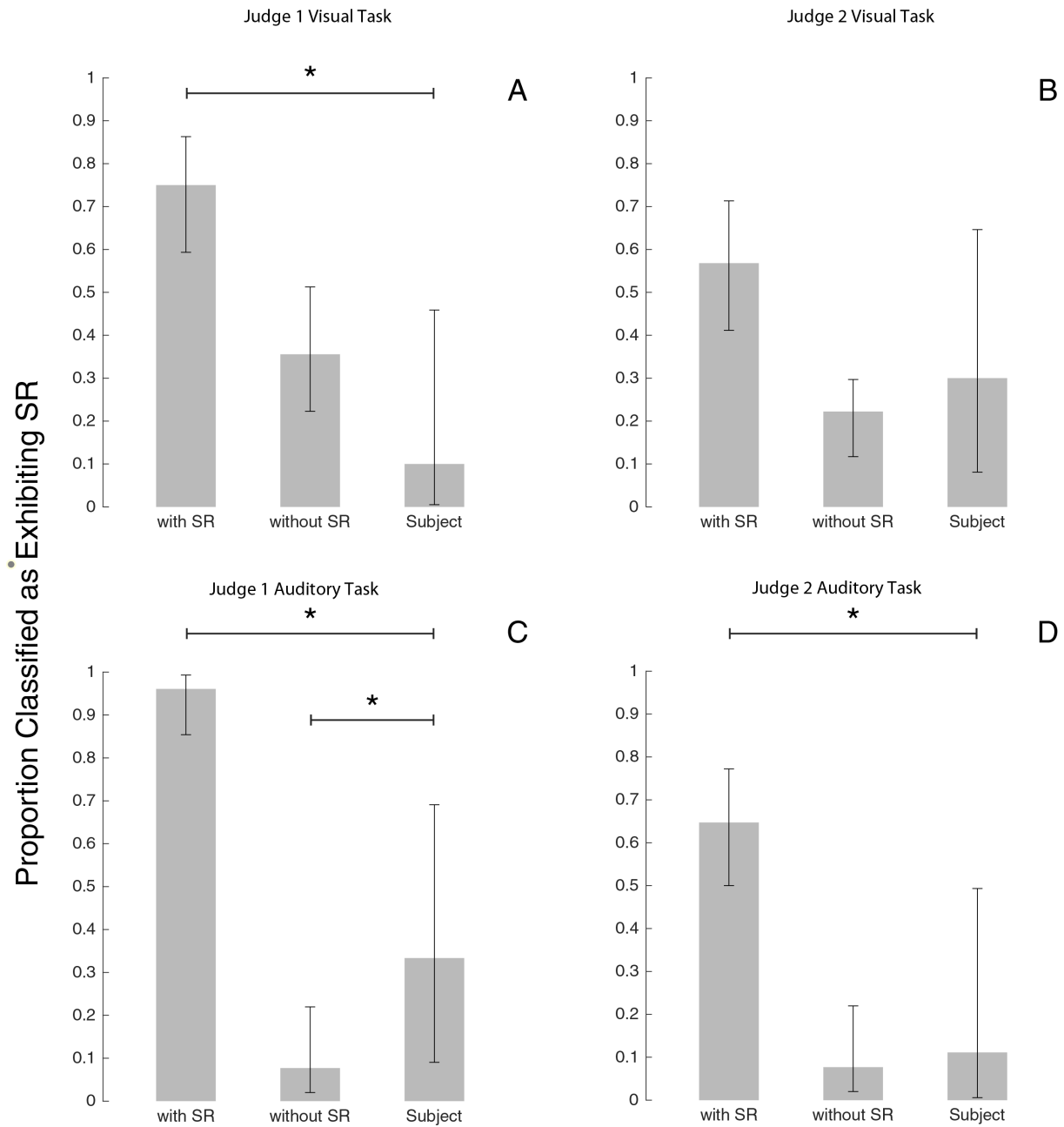

*Figure 6 Bar plots to show how the judges classified each group. Panel A: Judge #1 on visual task, panel B: Judge #2 on visual task, panel C: Judge #1 on auditory task and panel D: Judge #2 on auditory task. Stars indicate a significant difference between classification proportions by Chi-square tests (see text for details).*

## Relationship Between Sham Threshold and bnGVS Improvement

Next, we examined the relationship between amount of perceptual improvement and sham threshold. Amount of improvement was defined as the difference between sham threshold and the bnGVS stimulated threshold, when re-measured (i.e., negative values correspond to improved thresholds). We found a significant negative correlation between sham threshold and improvement in visual contrast thresholds (Pearson correlation $r(7) = -0.83$, $p = 0.005$). Unsurprisingly (since auditory thresholds did not improve with the bnGVS level), no such correlation was found in auditory thresholds ($r(7) = -0.39$, $p = 0.3$).

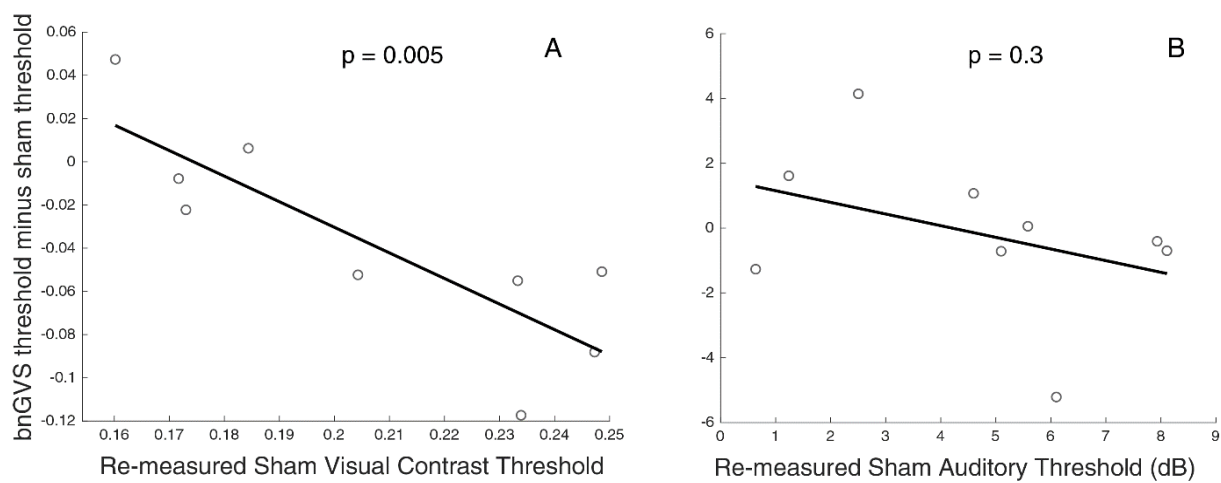

*Figure 7 Scatterplots of sham threshold against improvement (negative difference indicating improved threshold) with line of best fit. Panel A: Visual contrast thresholds. Panel B: Auditory thresholds. Correlation between improvement and sham threshold was only found in visual contrast thresholds.*

# Discussion

We have designed and implemented a statistically rigorous method of identifying cross-modal improvements in auditory and visual perceptual thresholds via the use of galvanic vestibular white noise stimulation. Our results demonstrate a statistically significant difference between sham and the subject-specific best nGVS level using independent samples, indicating that the addition of low levels of vestibular white noise elicits improvement in visual contrast thresholds.

Cross-modal improvement in visual thresholds is consistent with previous findings that used auditory white noise [8], [9]. Crucially, we have provided further evidence that cross-modal SR exists in human sensory perceptual thresholds using a new modality (vestibular stimulation) and in a more rigorous manner. Through the re-measurement procedure, we ensured independent samples on which to run a statistical test. This is an improvement upon previous studies, which either did not perform any statistical test [8] or a re-measurement and thus producing sampling bias in the threshold measurement at the "best" noise level [9].

We found a negative correlation between baseline (sham) threshold and improvement in visual perception. Specifically, we found that those with worse visual contrast thresholds stood to benefit the most from nGVS. Galvan-Garza et al. [6] found a similar relationship for in-channel vestibular roll tilt perceptual thresholds. If individuals with innately higher thresholds are the most susceptible for enhancement, there may be benefits of GVS white noise for patient populations.

While we found GVS white noise improved visual thresholds, it did not significantly change auditory thresholds. When SR-benefits are not observed, there are multiple speculative explanations. It may be that a different auditory tone duration (other than 0.25 seconds) would be more conducive to cross-modal SR (visual presentations were 1 second). Although [11] found in-channel auditory SR at 1kHz, it is possible that the same 1kHz frequency might not be conducive to cross-modal SR. Alternatively, a different range of GVS white noise levels, profile, or application procedure may be necessary. Further research is needed to determine if indeed GVS white noise is ineffective at producing SR-benefits in auditory perception, but our results support the null hypothesis that nGVS does not affect auditory thresholds.

We found there was not one GVS level (or small range of GVS levels) that produced the lowest thresholds for all or even most subjects. This has not be systematically assessed for cross-modal SR.

For in-channel SR, Galvan-Garza et al. [6] found vestibular perceptual roll tilt thresholds were significantly improved across all subjects at 0.3 and 0.5 mA (but not at 0.2 and 0.7 mA, the other levels assessed), suggesting some amount of consistency in each subject's best nGVS level. Alternatively, Keywan et al. [7] found the best nGVS level varied between individuals fairly substantially (0.05 to 0.3 mA, mean = 0.135±0.86 mA, when testing at 0.05, 0.1, 0.15, 0.2, 0.3, 0.4, and 0.5 mA), as identified using a balance task. It should be noted that our study and these other two studies used slightly different protocols for applying nGVS, such that amplitudes should not be compared directly across studies. Instead, we conclude that while in-channel vestibular SR may benefit most subjects using a single nGVS level [6], for the cross-modal benefits to visual perception we observed its critical to identify subject-specific best nGVS levels.

We have not yet shown that the improvement is consistent with existing SR models, as has been shown for in-channel vestibular stimulation [6]. Higher plot classification accuracy has potential to generate more conclusive results with respect to SR identification. Notably, when judge #1 performed with very high accuracy while classifying auditory task data (Figure 6C), it became much easier to identify differences between the subject pool and simulated conditions. We speculate that more accurate and objective plot classification may be possible with algorithmic classification (instead of using human judges). Additionally, it is possible that an underlying curve with a smaller threshold improvement with nGVS level would be more representative. In particular, judge #1 on the auditory task classified subjects differently from both groups of simulations. This indicates that perhaps the subject group did exhibit SR-behaviour in auditory thresholds with nGVS, but that the underlying model we used for comparison had too great of a threshold improvement. Regardless of these limitations, these classification methods are the currently best practices and enhance rigor aimed at identifying a characteristic u-shaped SR response.

Our study was scoped to identify cross-modal benefits of nGVS, but was not scoped to investigate potential mechanisms, so instead we briefly speculate how nGVS could improve visual thresholds. Multisensory neurons have been shown to exist in both animals and humans [41]–[43] and cross-modal SR is thought to use them [8]. There are currently several models for how multisensory information is processed [44]. Some models use a linear combination of cues [45], [46], while others use probabilistic inference [47], [48] based on reliability of each sensory cue. One study that examined cross-modal SR in the auditory channel with tactile noise hypothesized that the occurrence of cross-modal SR in that modality may be due to the dorsal cochlear nucleus which combines both auditory and somatosensory cues [32]. It is possible that a requisite nucleus exists for the visual and vestibular systems. Based on current models of multisensory perception, two sensory cues occurring at the same time (e.g., visual stimulus and nGVS) in integrated sensory channels (such as visual and vestibular) may be important for the mechanism of cross-modal SR improving perception of the stimulus.

## Conclusions

We conclude that galvanic vestibular white noise stimulation results in cross-modal improvements in the visual channel in that it lowers visual contrast thresholds. We found a correlation between subjects' sham threshold and their improvement magnitude. Future research is necessary to identify the mechanism behind the cross-modal improvement and to appropriately model the reduction in perceptual thresholds.

Auditory thresholds appear similar with and without vestibular white noise stimulation. Should improvement in auditory thresholds exist with vestibular white noise stimulation, the improvement may not be large enough to be captured by our study size or threshold measurement precision.

## Declaration of Interest

## Acknowledgements

This study was funded by NASA TRISH under award number T0402. This study was part of a master's thesis [49].